\documentclass[aps,prb,preprint,groupedaddress,superscriptaddress,showpacs]{revtex4}
\usepackage{graphicx, bm}
\usepackage[usenames]{color}
\usepackage{soul}

\begin{document}

\title{Organic Magnetoelectroluminescence for Room Temperature Transduction between Magnetic and Optical Information}
\author{Ferran Maci\`a$^*$}
\affiliation{Department of Physics, New York University, 4 Washington Place, New York, New York 10003, USA}
\author{Fujian Wang$^*$}
\author{Nicholas J. Harmon}
\affiliation{Department of Physics and Astronomy and Optical Science and Technology Center, University of Iowa, Iowa City, Iowa 52242, USA}
\author{Andrew D. Kent}
\affiliation{Department of Physics, New York University, 4 Washington Place, New York, New York 10003, USA}
\author{Markus Wohlgenannt}
\author{Michael E. Flatt\'e}
\affiliation{Department of Physics and Astronomy and Optical Science and Technology Center, University of Iowa, Iowa City, Iowa 52242, USA}

\date{\today}

\begin{abstract}
Magnetic and spin-based technologies for data storage and processing pose unique challenges for information transduction to light because of magnetic metals' optical loss, and the inefficiency and resistivity of semiconductor spin-based emitters at room temperature\cite{Awschalom:2007:NatPhys}. Transduction between magnetic and optical information in typical organic semiconductors poses additional challenges as the Faraday and Kerr magnetooptical effects rely on the electronic spin-orbit interaction\cite{Meier1984}, and the spin-orbit interaction in organics is weak\cite{Naber:2007:JPhysD,Vardeny2010}. Other methods of coupling light and spin have emerged in organics, however, as the spin-dependent character of exciton recombination,  with spin injection from magnetic electrodes, provides magnetization-sensitive light emission\cite{Dediu2002,Davis2003}, although such approaches have been limited to low temperature and low polarization efficiency\cite{Nguyen:2012:Science}. Here we demonstrate room temperature information transduction between a magnet and an organic light emitting diode that does not require electrical current, based on control via the magnet's remanent field of the exciton recombination process in the organic semiconductor.
\end{abstract}


\maketitle
Organic semiconductor devices have become a ubiquitous lighting technology due to their flexibility, inexpensive character, and highly efficient light output. Organic information processing also has an important niche where computational speed is less important than flexibility or expense. Nonvolatile information storage in organic electronics poses a challenge, as organic semiconductor  flash memory currently have high leakage currents due to the large electric fields required to change transport properties in organic semiconductors\cite{Sekitani2009}. The non-volatility, speed, and low energy consumption of magnetic memory make it fundamentally attractive for integration with organics although magnetic metals have large impedence mismatches with organics. This can be overcome using coupling through a magnet's fringe field at zero applied field (remanent field) as demonstrated in organic electronic devices\cite{Wang:2012:PRX,Macia:2013:APL, Harmon:2013:PRB}.
Efficient coupling via the magnet's  remanent field  between light emission in organic devices and a magnetic memory would broaden the range of applicability of flexible, inexpensive organic electronics. Our demonstration of efficient room-temperature coupling between an organic light-emitting diode and a few-nanometer-thick magnetic film can be explained quantitatively within a theory of spin-dependent exciton recombination in the organic semiconductor, influenced only by the remanent fringe fields of the magnetic material.

\begin{figure}
 \includegraphics[width=\columnwidth]{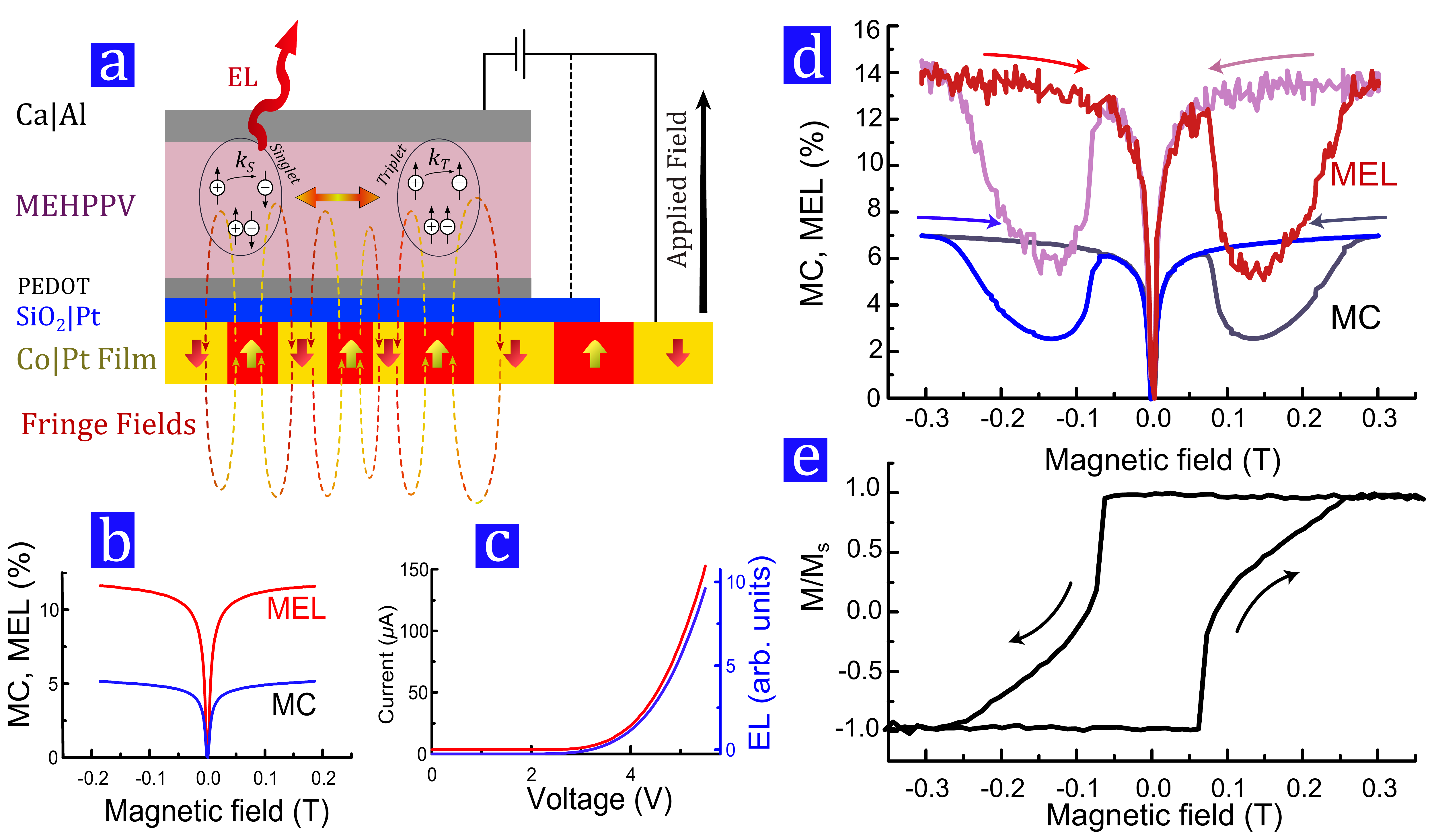}
  \caption{\small{\textbf{Magnetoconductance (MC) and magnetoelectroluminescence (MEL) in organic semiconductor devices:} \textbf{a}. Schematic of the device structure used for fringe-field MC and MEL measurements. The device consists of a standard OMAR/OLED device fabricated on top of a ferromagnetic film, which need not be in electrical contact with the organic device. A SiO$_{\rm 2}$ layer was used in some devices to electrically isolate the magnetic film from the organic device. The electroluminescence (EL) is collected through the semitransparent top contact. \textbf{b}. MC and MEL responses to an external magnetic field of a \emph{reference} OMAR/OLED device without the magnetic film, and \textbf{c} $IV$ and $EL$ versus voltage for the reference device. \textbf{d}. MC and MEL responses of the complete organic fringe-field device. \textbf{e}. Magnetization $M$ relative to the saturation magnitization $M_S$ of the ferromagnetic film as obtained by MOKE. All data are for room temperature. \label{fig:MainExperimentalFigure}}}
\end{figure}

Organic semiconductor sandwich devices, used for example in organic light-emitting diodes (OLEDs), consist of a thin film of an organic semiconductor  (or several layers thereof) sandwiched between a bottom and top electrode. The organic semiconductor is typically intrinsic, and is essentially void of charge carriers. Therefore one of the electrodes is chosen to efficiently inject electrons, and the other to efficiently inject holes. For this purpose low and high-work function metals, respectively, are chosen. Recombination of electron-hole pairs leads to electroluminescence, as shown in Fig.~\ref{fig:MainExperimentalFigure}a. Transport of the injected carriers through the organic film occurs via a sequence of hops along a path connecting the top electrode to the bottom electrode, and the rate of transport and/or recombination is dramatically affected by variations in the local magnetic field along the path, as found in numerous experimental and theoretical studies \cite{Kalinowski:2003:ChemPhysLett,Francis:2004:NewJPhys,Prigodin:2006:SynthMet,Bobbert:2007:PRL,Desai:2007:PRB,Hu:2007:NatMat}. In the transport regime, this effect is known as organic magnetoresistance (OMAR), and the corresponding effect in the electroluminescent output we will denote as organic magnetoelectroluminescence (OMEL). Typically the source of that inhomogeneous field is the nuclear hyperfine field, which is random and spatially uncorrelated. The origin of OMAR can be traced back to interactions between paramagnetic charge pairs that occur at bottle-neck sites (sites that crucially affect the transport and/or electroluminescent properties). These interactions are often spin-selective, and the reaction rate depends on the angle between the local spin-quantization axis (provided by the local magnetic field) for the two paramagnetic species. An applied magnetic field exceeding the hyperfine field strength forces this angle to be close to zero everywhere in the film, which shows up as magnetoconductance (MC) and/or magnetoelectroluminescence (MEL) with typical features like those in Fig.~\ref{fig:MainExperimentalFigure}b.

Large gradients in the magnitude of the random local field can lead to fringe-field-driven MC and MEL, and devices were constructed that utilize fringe fields from an unsaturated magnetic film to supply such gradients \cite{Wang:2012:PRX,Macia:2013:APL, Harmon:2013:PRB}. This paper will primarily discuss such fringe-field effects, but ``ordinary'' OMAR and OMEL (based on, {\it e.g.}, hyperfine fields) is also investigated as a reference. The exact nature of the paramagnetic pairs remains hotly debated, and the possibilities include electron-hole pair recombination to form singlet or triplet excitons \cite{Prigodin:2006:SynthMet}, $e-e$ or $h-h$ recombination to form singlet bipolarons \cite{Bobbert:2007:PRL}, and collision reactions between electrons or holes with long lived triplet excitons \cite{Desai:2007:PRB}. As the fringe-field MEL mechanism is driven by fringe-field gradients, independent of the mechanism of ordinary OMAR and OMEL, the effects presented here should be found in organic materials independent of the mechanism of OMAR and OMEL, including in materials where there is no OMAR and OMEL\cite{Harmon:2013:PRB}.

As shown in Fig.~\ref{fig:MainExperimentalFigure}\textbf{a} our devices are OMAR/OLED devices fabricated on top of a ferromagnetic thin film. The device fabrication starts with the metal deposition of a ferromagnetic multilayered thin film made of Cobalt (Co) and Platinum (Pt). Those films have perpendicular magnetic anisotropy; the spins tend to align in the direction orthogonal to the film plane. In presence of a large magnetic field out of the film plane the Co$|$Pt films are uniformly magnetized with all the spins pointing opposite to the direction of the applied field. At lower fields the films form magnetic domains---some regions with spins pointing up and others with spins pointing down---to lower the magnetostatic energy. These magnetic domains create strong varying fringe fields close to the surface of the Co$|$Pt films, which penetrate the OMAR/OLED device. In order to electrically insulate the OMAR/OLED device from the ferromagnetic film we deposit a thin dielectric followed by a conductive nonmagnetic layer on top of the magnetic film. These electrically insulated magnetic films prove that fringe fields---and not electrical currents---are responsible for the coupling between the ferromagnetic layer and the OMAR/OLED device. However, the strength and spatial correlation length of magnetic fringe fields depend sensitively on the distance from the magnetic film to the organic film. The insulating layer increases this distance reducing the effect. As a result we report results primarily on devices without this layer, which show the largest effects, and study devices with insulating layers to rule out spin injection and tunneling anisotropic magnetoresistive devices.

A conducting polymer layer (20~nm) of poly(3,4-ethylenedioxythiophene) poly(styrenesulfonate) (PEDOT:PSS, referred to as simply PEDOT from now on) was spin-coated from water suspension and serves as the hole-injecting layer. PEDOT is commonly employed for this purpose in organic light-emitting diode devices.\cite{Kim:2002:APL} We chose poly[2-methoxy-5-(2-ethylhexyloxy)-1,4-phenylenevinylene] (MEHPPV) as the luminescent polymer, as it is widely used as a red emitter in OLEDs\cite{Malliaras:1998:PRB}. The MEHPPV layer (55~nm) was deposited by spin-coating from toluene solution. Finally, Calcium (Ca) (6 nm, serving as the electron-injecting top contact) covered by Aluminum (Al) (12~nm) was deposited by vacuum evaporation through a shadow mask. The active device area is roughly 1~mm$^2$. The Al capping layer is required to protect the highly reactive Ca layer. The contributions of the PEDOT and Ca electrodes to the device resistance (and magnetoresistance) are negligible, since they are metals, whereas MEHPPV is an intrinsic semiconductor. The EL is measured through the semitransparent top electrode, and recorded by a photomultiplier tube. All measurements reported here are at room temperature.

Figure~\ref{fig:MainExperimentalFigure}\textbf{b} shows a typical MC and MEL trace, and Fig.~\ref{fig:MainExperimentalFigure}\textbf{c} the IV and EL curves, for an organic device without a magnetic film and whose MC/MEL is therefore caused by the random hyperfine fields, as described above. This device will serve us as a reference when, later on, we will discuss fringe-field induced MC/MEL. It is seen that the hyperfine induced MC and MEL responses have a magnitude of $\approx$ 5 \% and $\approx$ 10 \%, respectively, in our MEHPPV devices. The effects essentially saturate for applied fields in excess of 0.1 T, are non-hysteretic, and have a full-width-at-half-maximum of approximately 20 mT. The effects are also independent of the direction of the applied magnetic field, and nearly independent of the MEHPPV layer thickness. In the present work, we have chosen to work with a thin MEHPPV layer (55nm) such that the distance from the ferromagnetic film does not vary much between different locations in the MEHPPV film.

Now we turn our attention to the MC/MEL responses of the fringe-field OMAR devices, and the correlation between these effects and the film magnetization, $M$. Figure~\ref{fig:MainExperimentalFigure}\textbf{d} shows the measured MC and MEL curves, and \textbf{e} shows the magnetization loop measured by magneto-optic Kerr effect (MOKE) (see methods section). In these measurements, the magnetic field is applied perpendicularly to the device plane, and is swept smoothly from large negative to large positive fields (black lines) and back (red lines). It is seen that the magnetization response is hysteretic, and that $M$ assumes its saturation value $M_S$ for fields larger than approximately 0.25~T in magnitude. $M$ is unsaturated between roughly $0.05$ and $0.25$~T. The MC/MEL curves outside the unsaturated magnetization regime clearly mirror the data in non-magnetic devices (panel \textbf{b}), and are explained by the ``normal'' hyperfine OMAR effect. In the unsaturated region, the data curves develop characteristic ``ears''. These are the signature of  fringe-field effects. We have previously given a detailed experimental and theoretical characterization of the transport aspect of this effect.\cite{Wang:2012:PRX,Macia:2013:APL, Harmon:2013:PRB} In the present work, we demonstrate for the first time that fringe-field effects lead to a sizable room-temperature MEL response, of up to 6\% at room temperature for the present device. This can be comparable to MEL effects that occur only at low temperature, such as those recently reported in spin-valves\cite{Nguyen:2012:Science} and high-magnetic-field effects for OLEDs\cite{Wang:2012:NatCommun}.

\begin{figure}[htp!!]
\centering
\includegraphics[width=0.75\columnwidth]{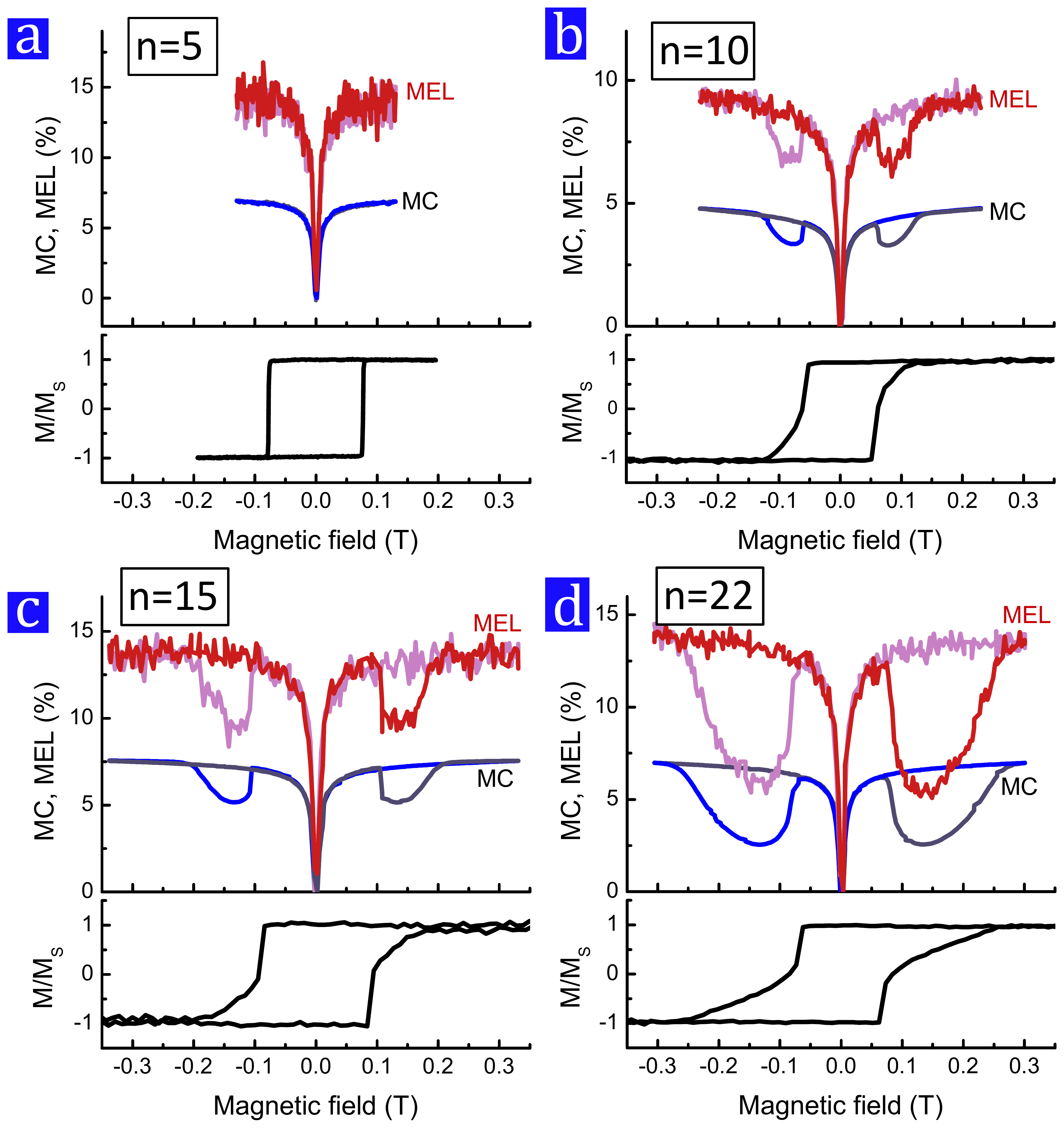}
\caption{\small{\textbf{Correlation between magnetoconductivity/magnetoelectroluminescence and magnetization loop for several different ferromagnetic films:}} \textbf{a-d.} MC and MEL responses of organic fringe-field devices using different ferromagnetic films. The magnetic films are Co$|$Pt multilayers with a different number of Co$|$Pt repeats, $n$, resulting in different magnetization loops as evidenced by the MOKE $M/M_S$. In all cases \textbf{a-d} the fringe-field response correlates exactly with the field-range where the magnetic film is unsaturated, and therefore emits fringe fields into the organic device.}
\label{fig:DependenceOnRepeats}
\end{figure}

Next we examine the relation between the magnetic film's response characteristics and MC/MEL by fabricating OMAR/OLED devices on several different ferromagnetic electrodes. In this work, the different magnetic responses are studied by fabricating ferromagnetic films consisting of Cobalt (Co) and Platinum (Pt) multilayers with a different number of repeats, $n$. We studied devices with $n=5,$ $10$, $20$ and 30 (film thicknesses varied from 4~nm to 24~nm). The magnetization in the ferromagnetic films reverses through nucleation, growth, and annihilation of magnetic
domains. When the magnetic films are saturated (all spins pointing towards the same direction) there are no magnetic fringe fields on top. When the magnetic films are unsaturated the strength of the magnetic fringe fields created by magnetic domains increases (almost linearly) with the thickness of the ferromagnetic layer (i.e., with the number of repeats). Properties of Co$|$Pt ferromagnetic films have been characterized in detail (see Refs. \onlinecite{Wang:2012:PRX} and \onlinecite{Macia:2013:APL}).
The data of Fig.~\ref{fig:DependenceOnRepeats} shows that the hysteretic magnetoresistance of the organic layer is directly correlated with the hysteretic magnetization of the ferromagnetic film. The fringe-field ``ears'' occur only in the unsaturated regime, where film's magnetization breaks into domains with fringe fields occurring near the domain boundaries.

\begin{figure}[htp!!]
\centering
\includegraphics[width=0.5\columnwidth]{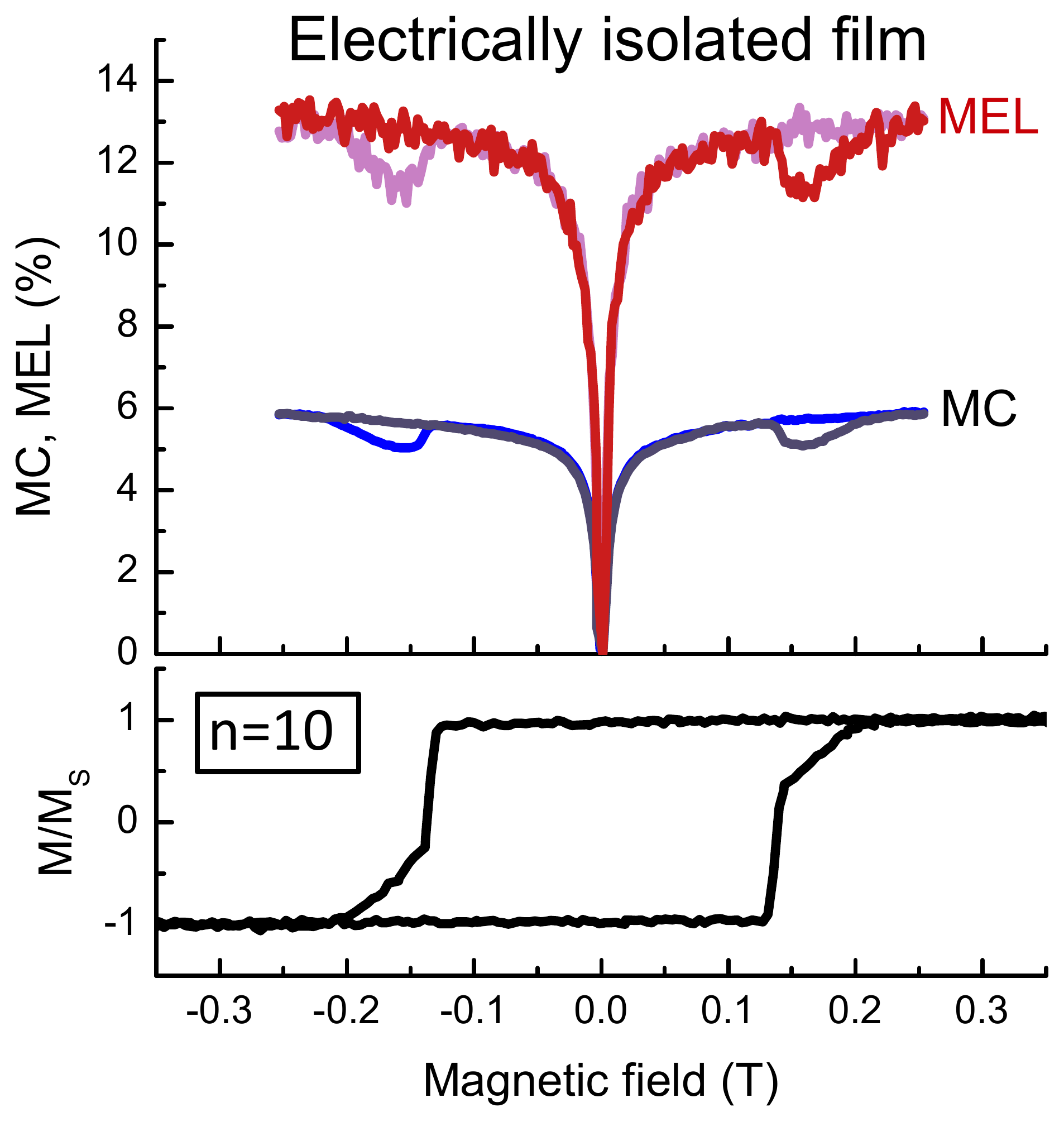}
\caption{\small{\textbf{Demonstration that the fringe-field effect is a result of magnetic, rather than electrical, coupling between the magnetic layer and the OMAR/OLED device:}} MC and MEL responses of an organic fringe-field device where the magnetic film is electrically isolated from the organic device by insertion of a SiO$_{\rm 2}$ layer between them. This control experiment is important for establishing the correct interpretation of the data (see text), but results in a greater distance between magnetic film and organic device, leading to a smaller coupling between the two and a smaller fringe-field effect.}
\label{fig:IsolatedDevice}
\end{figure}

Figure~\ref{fig:IsolatedDevice} shows data similar to that reported in Fig.~\ref{fig:DependenceOnRepeats}, but now for a device with an additional SiO$_{\rm 2}$ layer inserted between the magnetic film and the OMAR/OLED device. This data exhibits all the same characteristics of the data without the insulating layer, and proves that the coupling between the magnetic film and the organic device is magnetic rather than electrical in nature. In particular this excludes mechanisms such as tunneling anisotropic magnetoresistance and spin-injection effects as the origin of the observed effects. The ``ears'' are however significantly smaller in magnitude. However, this was to be expected, since the insulating layer leads to a significantly larger separation between magnetic film and OMAR/OLED device as the overall spacer layer thickness increases from 20 nm to 45 nm with the layer inserted. We have previously examined\cite{Macia:2013:APL} the dependence of the magnitude of the fringe-field magnetoresistance on the spacer layer thickness.

\begin{figure}[htp!!]
\centering
\includegraphics[width=0.7\columnwidth]{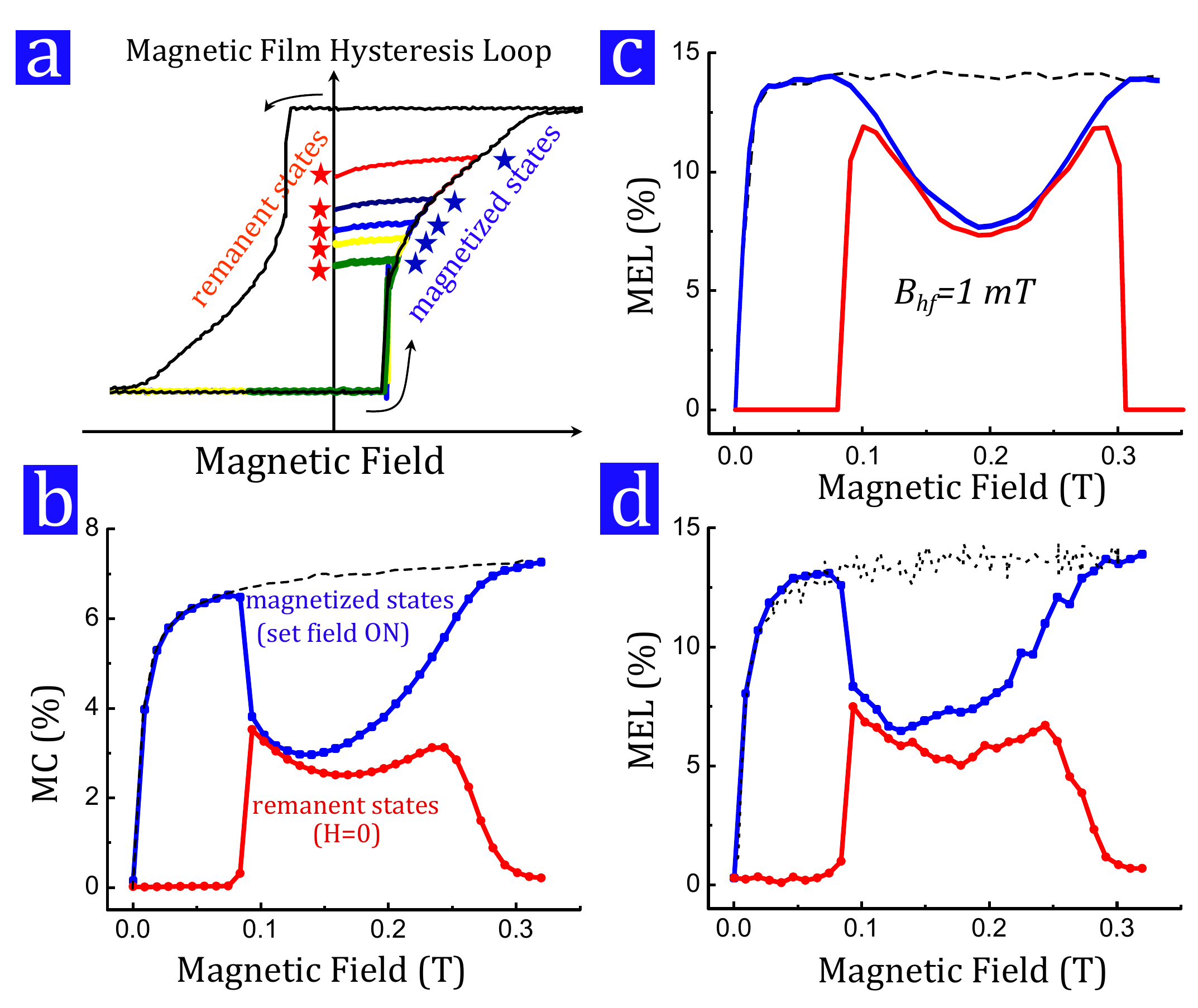}
\caption{\small{\textbf{Correlation between remanent magnetic state of the magnetic film and the conductivity/electroluminescence measured in the organic semiconductor device:} In these measurements the applied magnetic field plays an auxiliary role, and is used to write (``set'') a particular remanent magnetic state, corresponding to a particular domain configuration and particular fringe-field pattern, and to erase (``reset'') this configuration by replacing it with a saturated state without domains. This saturated state without domains is also used as the reference state against which we measure the MC and MEL percentage responses. Referring to the magnetization loop of an example magnetic film, \textbf{a}} shows the procedure to set and reset different remanent states: The set field (blue star) is used to select a magnetic state. As the set field is being removed, the film remains in essentially the same state as evidenced by the only small amount of relaxation in magnetization between write state (blue star) and remanent state (red star).
\textbf{b} Conductance  and \textbf{d} electroluminescence referenced to a fully saturated state measured in different magnetic states of the ferromagnetic film vs. set field strength (red curve). For comparison, the conductance and electroluminescence value measured while the set field is still on is also shown (blue curve). Grey dashed curves correspond to measured values of conductance (\textbf{b}) and electroluminescence (\textbf{d}) at the set field when the magnetic film is saturated (blue and purple curves in Fig.~\ref{fig:DependenceOnRepeats}). \textbf{c} Theoretical calculation following text, using switching fields from Fig.~\ref{fig:MainExperimentalFigure}\textbf{e}, to be compared with \textbf{d}. Grey dashed curve is calculation with fringe fields absent.}
\label{fig:RemanentMEL}
\end{figure}

Magnetic domains can be present at zero applied field in our magnetic films. Such remanent states are prepared by applying a perpendicular field close to the film's coercive field and then removing it (see Fig.~\ref{fig:RemanentMEL}\textbf{a}). Magnetization measurements and imaging prove that remanent domain states relax only slightly upon removal of the field. Therefore, we have access to remanent magnetization states ranging from negative to positive saturation. At zero applied field we observed how remanent fringe fields increase the conductance of the organic layer MEHPPV, suppressing OMAR. The same effect with a smaller strength was observed in the organic semiconductor Alq$_3$\cite{Macia:2013:APL}. Figure\ \ref{fig:RemanentMEL}\textbf{bd} shows MC and MEL of a 55~nm thick MEHPPV film on top of a magnetic layer both in presence of and at zero magnetic field. The blue line depicts the measured values in presence of magnetic field whereas the red lines trace the values measured after removing the applied field (a sketch of the measuring sequence is shown in Fig.~\ref{fig:RemanentMEL}\textbf{a}): We first saturated the sample with a large negative field, then we set a positive field value from 0 to 0.3~T and measured the conductivity of the organic layer (blue points), and then we removed the applied field and again measured the organic's conductivity (red points). Here we show that the electroluminiscence increases up to 6$\%$ for remanent magnetic domain states of the ferromagnetic layer. In contrast to fringe fields from the same domain configuration in an applied magnetic field (near the coercive field) the MEL increases rather than decreases.

We now examine whether the large fringe-field effects observed in the MEL can be explained by theory.
We consider a two-site model where an electron and hole (a polaron pair) occupy two nearby sites. The spin configuration of the polaron pair undergoes transitions due to the different magnetic interactions present; in our case these interactions consist of
\begin{equation}
H_0 = \omega_0 \hat{z} \cdot (\mathbf{S}_1 + \bm{S}_2), \ \ \ H_{hf} = \bm{\omega}_{hf_1} \cdot \mathbf{S}_1 + \bm{\omega}_{hf_2} \cdot \mathbf{S}_2,\ \ \ H_{ff} = \bm{\omega}_{ff_1} \cdot \mathbf{S}_1 + \bm{\omega}_{ff_2} \cdot \mathbf{S}_2
\end{equation}
which are the applied, hyperfine, and fringe-field Hamiltonians respectively. The polaron pairs recombine into excitons at different rates, $k_S$ and $k_T$, depending on the pair's spin since the singlet and triplet states have different energies and wavefunctions\cite{Kersten2011a} (see. schematic Fig.\ \ref{fig:MainExperimentalFigure}\textbf{a}); alternatively the pair could disassociate at a rate $k_D$. Once an exciton is formed, the large exchange energy precludes any further spin evolution. In the absence of large spin-orbit interactions, spin selection rules dictate that exciton recombination (i.e. photon emission) occurs only from the singlet exciton state. Assuming that radiative recombination is the only viable pathway for a newly formed singlet exciton, each singlet exciton will produce a single photon such that MEL can be defined in terms of $X_S$, the singlet fraction of excitons: $[X_S(B_0) - X_S(0)]/X_S(0)$.

To calculate the MEL we employ the stochastic Liouville equation for the polaron pair spin density matrix, $\rho$:\cite{Kubo1963, Haberkorn1976}
\begin{equation}
\frac{\partial \rho}{\partial t} =  - i [H_0 + H_{hf} + H_{ff},  \rho] - \frac{1}{2} \{  k_S P_S + k_T P_T, \rho \} - k_D \rho,
\end{equation}
where $P_S$ and $P_T$ are the singlet and triplet projection operators.
The steady-state singlet and triplet exciton fractions are\cite{Jones2010}
\begin{equation}
X_i = k_i \int_0^{\infty} \text{Tr}[P_i \rho(t)]dt,
\end{equation}
where $i$ runs over $S$ and $T$.
All rates (times) are in units of the hyperfine field frequency (period), $\gamma_e B_{hf}$ ($1/\gamma_e B_{hf}$) where $B_{hf}$ is the width of the Gaussian distribution of hyperfine fields. For simplicity we assume $\gamma_e = \gamma_h$ and $k_D = 0$.

To proceed with the calculation, one must have knowledge of the fringe fields present in the organic layer. In our previous analysis\cite{Harmon:2013:PRB} on fringe-field induced magneto-resistance, elementary magnetostatics were used to calculate fringe fields from XMCD images of the magnetic domains.
The samples described herein have the same composition. We use therefore the statistical analysis from the aforementioned XMCD images to model the fringe-field distributions.
Given the fact that fringe fields vanish at magnetic saturation and are largest and most varying at $M = 0$, we model the fringe-field distribution as normal distribution with mean zero and a field dependent standard deviation. Each component of the fringe-field gradient, $B_{ff}$ is modeled likewise. The field dependent standard deviations follow a parabola defined in the upper half-plane according to
\begin{equation}
\sigma_{B_{ff_i}}(B_0) =  - \frac{B_{ff_{max}}}{(B_L - B_C)^2} (B_0 - B_C)^2 + B_{ff_{max}},
\end{equation}
\begin{equation}
\sigma_{G_{ff_i}}(B_0) =  - \frac{G_{ff_{max}}}{(B_L - B_C)^2} (B_0 - B_C)^2 + G_{ff_{max}},
\end{equation}
where $B_L$ is the field at which the magnet starts developing domains (obtainable from either the MOKE data or the MEL measurements) and $B_C$ is the field corresponding roughly to $M = 0$. $B_{ff_{max}}$ and $G_{ff_{max}}$ mark the parabola's vertex and are the only free parameters. However, from analysis of samples of the same composition on which XMCD images have been acquired, their values are roughly 40 mT and 1 mT/nm respectively for a spacer width of 20 nm.

The results of this theory and model are shown in Fig.\  \ref{fig:RemanentMEL}\textbf{c}. The $B_L$ and $B_C$ are chosen from Figure \ref{fig:MainExperimentalFigure}\textbf{e}.
The values of $k_S$(=$1.5\gamma_eB_{hf}$) and $k_T$(=$5\gamma_eB_{hf}$) are chosen to achieve a saturated MEL percentage near that of Figure \ref{fig:MainExperimentalFigure}\textbf{d} ($\sim 14 \%$).
The values for the hopping rates fall into the intermediate hopping regime ($k_S \sim \gamma_e B_{hf}$) which is necessary to have any fringe-field effect.

Remanent fringe fields generated from the domain structure of the magnetic film cause the observed dramatic modification of the electroluminescence from an organic light emitting diode at room temperature. As a uniform, perpendicularly magnetized film produces no remanent fields, the source of these fringe fields is the regions where the magnetization changes most rapidly, corresponding to domain walls. The faster the magnetization changes, or the smaller the domain size, the larger the remanent fields. Thus this approach of interfacing magnetic information encoded in the domain structure with an organic light emitting diode should become more effective and efficient as the magnetic domain sizes shrink. We note as well that the MEL is  significantly larger than the MC, indicating that (for MEHPPV) fringe-field optical coupling and readout will have greater sensitivity than fringe-field electrical coupling and readout of magnetic information.

\section{Methods}
\subsection{Fringe-field device fabrication}

The organic semiconductor semi-spin valve consists of a ferromagnetic layer, a hole-injecting layer, an organic semiconductor, and a top electrode. The ferromagnetic electrode is a Co$|$Pt multilayer  with the number of repeats varied from 4 to 22  deposited using electron-beam evaporation in ultra high vacuum on oxidized Si wafers for device studies and Si supported Si$_{\rm 3}$N$_{\rm 4}$ membranes for magnetic domain imaging studies using an x-ray transmission microscope. Optical lithography is used to define lines in the ferromagnetic thin film. A hole-injecting layer, conducting polymer poly(3,4-ethylenedioxythiophene) poly(styrenesulfonate) (PEDOT:PSS), was deposited by spin-coating from an aqueous suspension (suspension purchased from H. C. Starck, CLEVIO P VP AI 4083). A 55~nm thick film of organic semiconductor MEHPPV (purchased from American Dye Source, Inc.) was deposited by spin coating with 3mg/ml solution in Toluene. The electron injecting layer, Ca (6~nm) covered by Al (12~nm) was deposited by thermal evaporation at room temperature through a metal stencil to obtain a cross point device geometry. The ferromagnetic electrodes were characterized by magnetic force microscopy (MFM), ferromagnetic resonance (FMR), vibrating sample magnetometery (VSM), and magneto-optical Kerr effect (MOKE).

\subsection{Measurements}

Magnetoresistance measurements were done in a closed-cycle He cryostat positioned between the poles of an electromagnet. The measurements reported here are all at room-temperature. Magnetoresistance (MR) measurements were performed using a Keithley 2400 sourcemeter. Electroluminescence was measured by photomultiplier tube through the top electrode Ca/Al. X-ray measurements were performed at the Advanced Light Source (ALS) at the Lawrence Berkeley National Laboratory. Images were taken with XM-1 zone-plate microscope at beamline 6.1.2.

\section{Acknowledgements\label{sec:acknowledgements}}

This work was supported by ARO MURI Grant No. W911NF-08-1-0317. FM thanks support from EU, MC-IOF 253214.

\end{document}